\documentclass[prb,preprint,showpacs,preprintnumbers,amsmath,amssymb]{revtex4}
\usepackage{graphicx} 
\usepackage{dcolumn} 
\usepackage{amsmath}

\begin{document}
\bibliographystyle{apsrev}

\draft
%

\title{Structural, electronic, and magneto-optical properties of YVO$_3$}

\author{A. A. Tsvetkov}
\altaffiliation{Present address: NSRIM Institute, University of Nijmegen, Toernooiveld 1, 6525 ED
Nijmegen, The Netherlands} \email{tsvetkov@sci.kun.nl}
\author{F. P. Mena}
\author{P. H. M. van Loosdrecht}
\author{D. van der Marel}
\affiliation{Materials Science Center, University of Groningen, Nijenborgh 4, 9747 AG Groningen,
The Netherlands}
\author{Y. Ren}
\affiliation{Experimental Facilities Division, Advanced Photon Source, Argonne National Laboratory,
Argonne, IL 60439}
\author{A. A. Nugroho}
\altaffiliation{On leave from Jurusan Fisika, Institut Teknologi Bandung, Jl. Ganesha 10 , Bandung
40132, Indonesia.}
\affiliation{Solid State Chemistry Lab, MSC, University of Groningen, Nijenborgh
4, 9747 AG Groningen, The Netherlands}
\author{A. A. Menovsky}
\affiliation{Van der Waals - Zeeman Institute, University of Amsterdam, Valckenierstraat 65, 1018
XE Amsterdam, The Netherlands}
\author{I. S. Elfimov}
\author{G. A. Sawatzky}
\affiliation{Department of Physics and Astronomy, The University of British Columbia, 334-6224
Agricultural Rd., Vancouver, B.C. V6T 1Z1 Canada}
\date{\today}
\begin{abstract}
Optical and magneto-optical properties of YVO$_3$ single crystal were studied in FIR, visible, and
UV regions. Two structural phase transitions at 75 K and 200 K were observed and established to be
of the first and second order, respectively. The lattice has an orthorhombic $Pbnm$ symmetry both
above 200 K as well as below 75 K, and is found to be dimerized monoclinic $Pb11$ in between. We
identify YVO$_3$ as a Mott-Hubbard insulator with the optical gap of 1.6 eV. The electronic
excitations in the visible spectrum are determined by three $d$-bands at 1.8, 2.4, and 3.3 eV,
followed by the charge-transfer transitions at about 4 eV. The observed structure is in good
agreement with LSDA+$U$ band structure calculations. By using ligand field considerations, we
assigned these bands to the transitions to the $^4A_{2g}$, $^2E_{g} + ^2T_{1g}$, and $^2T_{2g}$
states. The strong temperature dependence of these bands is in agreement with the formation of
orbital order. Despite the small net magnetic moment of 0.01 $\mu_B$ per vanadium, the Kerr effect
of the order of $0.01^\circ$ was observed for all three $d$-bands in the magnetically ordered phase
$T_{\text{N\'eel}}<116 K$. A surprisingly strong enhancement of the Kerr effect was found below 75
K, reaching a maximum of $0.1^\circ$. The effect is ascribed to the non-vanishing net orbital
magnetic moment.
\end{abstract}

\maketitle

\section{Introduction}

The interplay between kinetic energy and Coulomb interaction in strongly correlated electron
systems enables small external or internal perturbations to manipulate electronic and magnetic
properties of the matter. If, in addition, there is a ground state degeneracy, further interesting
phenomena, like an orbital ordering can arise\cite{KugelKhomskii}.

In yttrium orthovanadate, YVO$_3$, spin and orbital ordering was suggested to be responsible for at
least one of the temperature induced sign reversals of magnetization\cite{Ren-Nature-1998}. In the
magnetically ordered phase, YVO$_3$ is a canted antiferromagnet, where a net ferromagnetic moment
is formed by a small angle between the antiferromagnetically (AFM) oriented spins of V$^{3+}$ ions.
The transition from the room temperature paramagnetic phase to the antiferromagnetic state occurs
at the N\'eel temperature $T_N=116$ K. Upon further lowering the temperature, the magnetic moment
gradually changes sign around 90 K. It has been argued that this magnetization reversal is due to
the opposing effects of the Dzyaloshinsky-Moriya interaction and the single-ion magnetic anisotropy
on the spin canting direction\cite{Ren-Nature-1998}. The magnetization switches sign again around
77 K. At this temperature the antiferromagnetic order changes from the high temperature $C$ type to
the low temperature $G$ type\cite{Kawano}. A $C$-type AFM order corresponds to an AFM arrangement
in planes and a ferromagnetic arrangement between the planes. A $G$-type AFM order implies
completely antiferromagnetic arrangement, both within planes and between them.  Ren {\it et.
al.}\cite{Ren-Nature-1998} argued that along with the spin order there is an orbital order on the
vanadium sites. According to the proposed picture, one of two $3d$ electrons always occupies the
$xy$ orbital, while another electron occupies alternatively either the $xz$ or $yz$ orbital on
different atoms, thus forming orbital order. In order to minimize the exchange energy a $C$-type
spin order is accompanied by a $G$-type orbital order and vise versa. The onset temperature of the
orbital order was reported to be much higher than the N\'eel temperature, $T_{OO}=200$
K\cite{Blake-PRB-2002}.

The proposed picture of the electronic structure still requires confirmation from, for instance,
spectroscopic experiments. The strength of optical transitions involving a charge transfer between
orbitals of different ions should be highly susceptible to orbital order between the ions. Indeed,
one of the aims of the present research is to study the influence of orbital ordering on optical
transition strengths, and use this technique to study orbital ordering in YVO$_3$. In this paper we
present, amongst others, optical and magneto-optical measurements of electronic excitations in
YVO$_3$ single crystals.  The observed electronic spectra are discussed using ligand field theory
considerations and compared to LDA+U calculations. In addition, magneto-optical Kerr effect
experiments are used to elucidate the spin orientation.

The crystal structure of yttrium orthovanadate YVO$_3$ was initially reported to have an
orthorhombic $Pbnm$ ($D_{2h}^{16}$ in Schoenfliess notation) symmetry at all
temperatures\cite{Rogers-JAP-1966,Zubkov-JETP-1974}. Recent synchrotron
measurements\cite{Blake-PRL-2001,Blake-PRB-2002} report the observation of an additional (401)
reflection below 200 K, inconsistent with the $Pbnm$ symmetry. The authors
concluded\cite{Blake-PRB-2002} that the crystal structure was most likely $P2_1/a$, the highest
symmetry subgroup of $Pbnm$. However, the intensity of the $Pbnm$ forbidden reflection was four
orders of magnitude weaker than to the allowed reflection\cite{Blake-PRB-2002}, which seemed to
make the refinement of the X-ray data very difficult. Far infrared (FIR) spectroscopy is a very
sensitive tool to study crystal structure variations. Minor atomic displacements due to a symmetry
breaking, which may be difficult to elicit from x-ray analysis, manifest itself through the
appearance of new phonon lines in the FIR spectrum. Recently, Ulrich {\it et.
al.}\cite{Ulrich-condmat-2002} proposed that YVO$_3$ can be subjected to a new sort of Peierls
instability, where the new ground state involves a dimerization along the $c$ axis. From the
crystallographic and optical points of view this dimerization removes the inversion center, and IR
and Raman active phonons become mixed. The $a$-axis modes stays nevertheless independent from the
$b$ or $c$ axes, if the dimerization affects only the $c$ axis. We use the FIR vibrational
spectroscopy to follow temperature induced variations of the phonon spectrum and to refine the
crystal structure of YVO$_3$.


\section{Experimental details}

Polycrystalline powder of YVO$_3$ was prepared from YVO$_4$ powder as starting material. The
YVO$_4$ powder was prepared by a high temperature solid-state reaction from appropriate mixtures of
pre-dried Y$_2$O$_3$ (99.998 \%) and V$_2$O$_5$ (99.995 \%, metal basis). The oxygen in YVO$_4$ was
reduced by annealing the powder in a flow of pure H$_2$ at 1000$^\circ$C. The growth process was
carried out by the floating zone technique using a four-mirror furnace in a flow of Ar
gas\cite{Ren-PRB-2000}. The single-crystalline boule of about 6 mm in diameter and 60 - 70 mm in
length was obtained from the growth. The crystallinity of the boule was checked by Laue X-ray
diffraction. The elemental composition of the crystal was checked by the electron probe micro
analyzer (EPMA) and the results showed that the molar ratios were given by Y:V:O = 1.00:1.00:3.02.
A separate check of the composition using the chemical analysis method showed that the cation ratio
of Y over V was 1.00 $\pm$ 0.01 and that the oxygen stoichiometry was 3.03 $\pm$ 0.02. Both results
were in good agreement.

The dielectric function for near-infrared, visible, and ultra-violent spectral ranges was obtained
directly by using the ellipsometry technique. Because ellipsometric measurements are very sensitive
to contamination of the surface, we used an ultra-high vacuum cryostat with a residual pressure
less than 10$^{-8}$ mbar. The cryostat was specially designed to minimize the movements of the
sample due to the thermal expansion of the cold finger. The spectra were measured with a VASE32
ellipsometer from J. A. Woollam Co., Inc., covering a spectral range from 6000 cm$^{-1}$ through
36000 cm$^{-1}$ (0.75-4.5 eV).

\begin{figure}
\includegraphics[clip=true,width=0.43\textwidth]{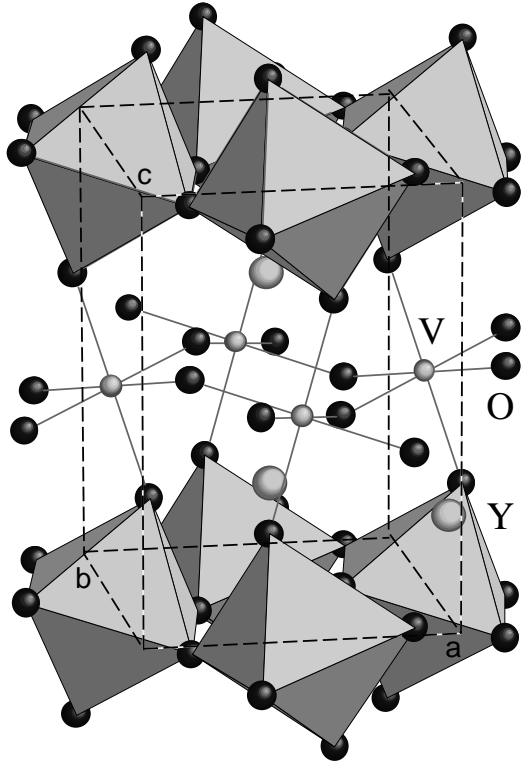}
\caption{\label{fig:Structure}$Pbnm$ orthorhombic crystal structure of YVO$_3$.}
\end{figure}

Far-infrared and mid-infrared reflectivity spectra were measured by Fourier Transform Infrared
reflection spectroscopy using a Bruker IFS-113v. The intensity of the reflected light from the
sample was referenced to that from a gold mirror for every measured optical surface at each
polarization. The sample was mounted on the cold finger of a helium flow cryostat, whose
temperature was varied from 300 K down to 4.2 K. Kramers-Kronig transformation was used to
calculate the dielectric function and optical conductivity from the reflectivity data in the
infrared region. For these calculations we extrapolated the reflectivity below 50 cm$^{-1}$ with a
pure ionic insulator response. For the high frequency extrapolation we used the reflectivity
calculated from the ellipsometric data.

The magneto-optical properties were measured using a home-built Kerr spectrometer. We used the
electro-optical modulation technique, similar to [\onlinecite{Sato-JJAP-1993}], to obtain
simultaneously the Kerr rotation and ellipticity induced by the sample. The combination of Xenon
arc lamp, windows, Si-detector, and other optical components limited our spectral range to the band
from 350 nm through 800 nm (1.55-3.5 eV).

\begin{figure}
\includegraphics[clip=true,width=0.47\textwidth]{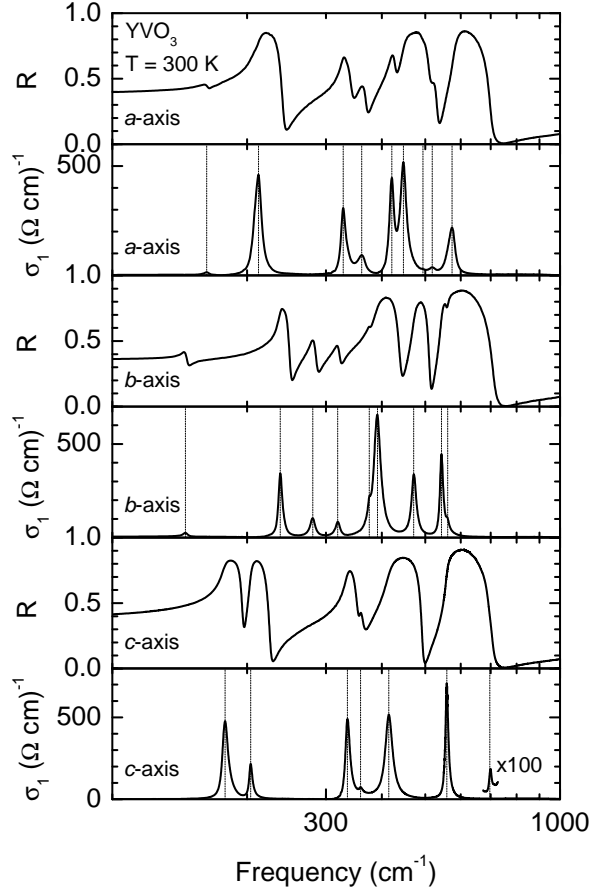}
\caption{\label{fig:ReflectivityABC}The room temperature reflectivity and corresponding optical
conductivity for the three crystallographic directions. Dashed lines indicate position of the
phonon modes. There are 9, 9, and 7 modes for the $a$, $b$, and $c$ axes, respectively, in
agreement with the $Pbnm$ symmetry.}
\end{figure}


\section{Structural phase transitions}

\begin{figure}
\includegraphics[clip=true,width=0.47\textwidth] {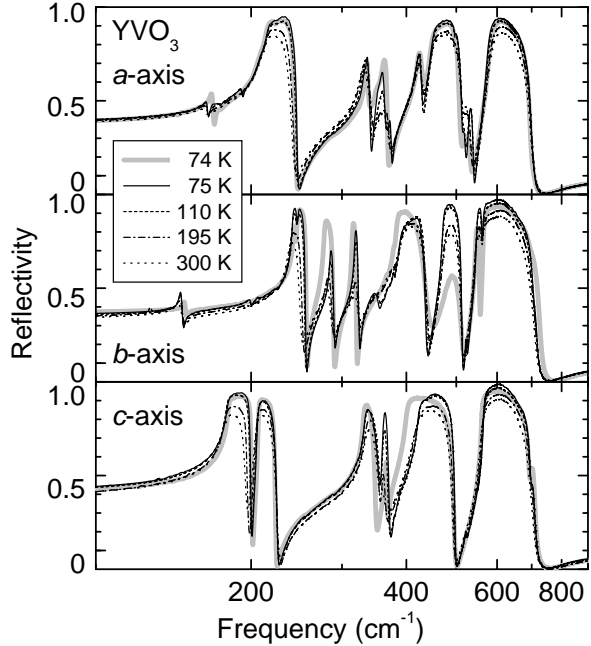}
\caption{\label{fig:Ref3polTdep} Temperature dependence of the FIR reflectivity for three
polarizations.}
\end{figure}

\begin{figure}
\includegraphics[clip=true,width=0.47\textwidth] {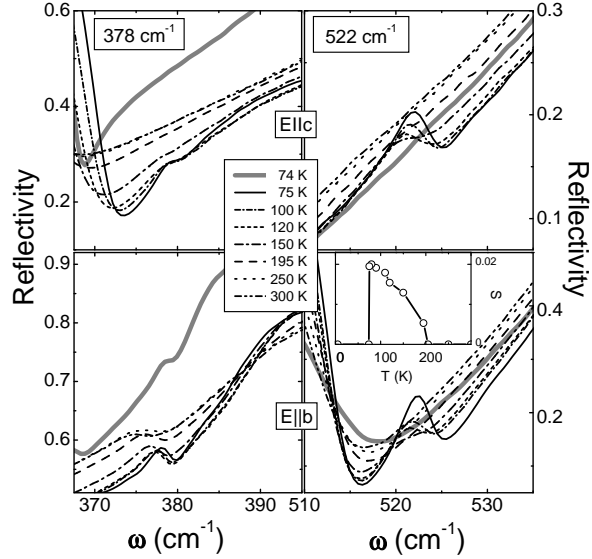}
\caption{\label{fig:NewModes} Left panels: ``Leakage'' of the b-axis $B_{3u}$ mode into the
$c$-axis response. Right panels: Appearance of a new previously silent $A_{u}$ mode in the $b$ and
$c$ responses. Insert: Temperature dependence of the optical strength of the new mode.}
\end{figure}

The orthorhombic $Pbnm$ crystal structure of YVO$_3$ is shown in Fig.~\ref{fig:Structure}. It can
be considered as a distorted perovskite structure. When compared to the perovskite unit cell, the
orthorhombic $a$ and $b$ axes lay along the diagonals of the perovskite unit cell and the
corresponding translational vectors are approximately $\sqrt{2}$ times longer. The $c$-axis
dimension of the orthorhombic unit cell is twice as large. The YVO$_3$ unit cell contains thus four
formula units. Vanadium ions have an octahedral surrounding. The octahedrons are distorted, rotated
and tilted with respect to each other, as illustrated in Fig.~\ref{fig:Structure}.

In the present work we use optically active phonons to track changes in the crystal symmetry.  As
yttrium orthovanadate is an insulating compound, its optical response in the far-infrared (FIR)
region is determined purely by lattice vibrations. In Fig.~\ref{fig:ReflectivityABC} the room
temperature reflectivity for all three polarizations is shown together with the optical
conductivity. As can be seen, there are 9, 9, and 7 optically active modes respectively for the
$a$, $b$, and $c$ axes. The positions of the TO phonon frequencies are shown by vertical lines and
the frequencies are given in Tables~\ref{table1} and \ref{table2}. Some resonances are very weak.
For example, the mode of highest frequency for the $c$ axis is hardly visible in the conductivity
and appears as a shoulder on the reflectivity curve. It gains intensity as the temperature is
lowered, and reaches its maximum in the low temperature orthorhombic phase. Shown in
Fig.~\ref{fig:Ref3polTdep} is the temperature dependence of the reflectivity for all three
polarizations. The number of the phonons stays the same between room temperature and 200 K. Below
200 K a number of additional phonons gradually emerges. We illustrate this by picturing some of the
new modes for the $b$ and $c$ axes in Fig.~\ref{fig:NewModes}. As can be seen from the figure, the
intensity of the new modes evolves continuously and monotonically as the temperature decreases,
reaching its maximum at 75 K. The optical strengths remain however one or two orders of magnitude
weaker than that of the original main modes. Below 75 K the new phonons discontinuously disappear.
As an example, the temperature dependence of the oscillator strength of the new 522 cm$^{-1}$ mode
is shown in Fig.~\ref{fig:NewModes}.

\begin{figure}
\includegraphics[clip=true,width=0.47\textwidth]{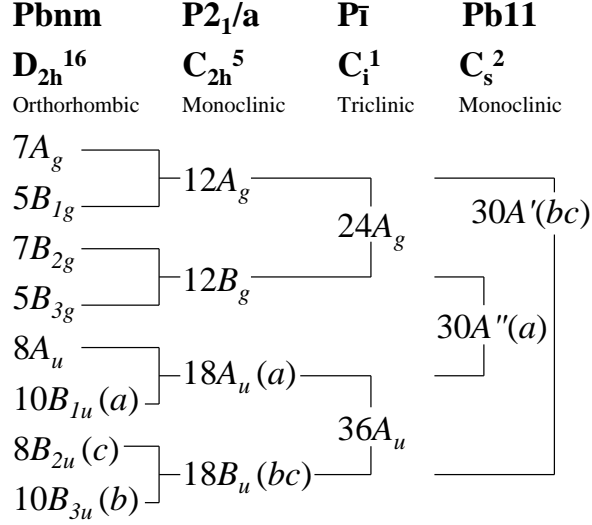}
\caption{\label{fig:DescentInSymmetry}Transformations of the phonon irreducible representations due
to the descent in lattice symmetry.}
\end{figure}

To analyze the observed phonon spectra, we apply a group theory analysis. The primitive cell of
YVO$_3$ contains four formula units, yielding 60 phonon modes in total. In the $Pbnm$ lattice the
four yttrium atoms and four vanadium atoms occupy equivalent sites 4$c$ and 4$b$ in Wyckoff
notation with the site symmetry $C_s^{xz}$ and $C_i$, respectively. The 12 oxygen atoms are
distributed between two inequivalent sites 4$c$ and 8$d$ with the site symmetry $C_s^{xz}$ and
$C_1$, respectively. Using the general procedure,\cite{Rousseau-JRS-1981} we find that the phonon
normal modes transform like 60 irreducible representations
\[
\Gamma = 7A_g + 8A_u + 5B_{1g} + 10B_{1u} + 7B_{2g} + 8B_{2u} + 5B_{3g} + 10B_{3u}.
\]
The total representation $\Gamma$ can be subdivided into three acoustic vibrational modes, $B_{1u}
+ B_{2u} + B_{3u}$, 24 Raman active modes, $7A_g + 5B_{1g} + 7B_{2g} + 5B_{3g}$, eight silent $A_u$
modes, and 25 infrared active modes. The infrared vibrations are split into $9B_{1u}(E||a)$,
$9B_{3u}(E||b)$, and $7B_{2u}(E||c)$ modes with the electric dipole moment aligned along the $a$,
$b$, and $c$ axes, respectively.

A similar group theory analysis is applied to the lower symmetry structures, $P2_1/a$,
$P\overline{1}$, and $Pb11$, which are sub-groups of the $Pbnm$ symmetry. Table~\ref{table3}
illustrates which symmetry elements are inherited in different symmetries. We also show a relation
between the widely used notation $Pbnm$, adopted also in this paper, and the standard notation
$Pnma$. The monoclinic $P2_1/a$ symmetry is postulated from the X-ray
measurements\cite{Blake-PRL-2001}. The triclinic $P\overline{1}$ symmetry, second highest possible
symmetry after $P2_1/a$, was proposed by us for the case, when the center of inversion is
conserved\cite{Tsvetkov-PhysicaB-2002}. Both x-ray\cite{Blake-PRL-2001} and optical
measurements\cite{Tsvetkov-PhysicaB-2002} show that the symmetry can not be higher than monoclinic.
Recent neutron experiments indicated the possibility of a dimerized
state\cite{Ulrich-condmat-2002}. Dimerization displaces the middle VO$_2$ plane, see
Fig.~\ref{fig:Structure}, away from its central position. It removes the inversion center and screw
axis from the symmetry operations of $P2_1/a$. The remaining symmetry group includes only a glide
plane $\mathbf{b}$ and is referred as $Pb11$.

As the X-ray measurements do not reveal any multiplication of the unit cell,\cite{Blake-PRB-2002}
the number of the atoms in the unit cell is not changed. For monoclinic $P2_1/a$ ($C_{2h}^5$)
symmetry group one expects $16B_{u}$ IR active and 2 acoustic modes in the $bc$ plane and $17A_u$
IR and one acoustic modes along the $a$ axis. In case of triclinic $P\overline{1}$ ($C_i^1$)
symmetry, there should be $33A_u$ IR modes, which can have components along all the three axes.
Monoclinic $Pb11$ ($C^2_s$) symmetry splits phonons into two irreducible representations
$30A'(E||bc) + 30A''(E||a)$, where Raman and IR modes are mixed. The way how the phonon modes'
irreducible representations transform into each other due to the descent in symmetry is depicted in
Fig.~\ref{fig:DescentInSymmetry}. Both monoclinic $C_{2h}^5$ and triclinic $C_i^1$ crystal lattices
have an inversion center. As a result the IR and Raman modes do not mix with each other for these
symmetries. In point group $C_{2h}$, because the $b$ and $c$ axes are not orthogonal any more, the
corresponding phonons mix and form the $B_u$ representation, $7B_{2u}+9B_{3u}\rightarrow 16B_{u}$.
The mode $A_u$, silent in $D_{2h}^{16}$ symmetry, becomes optically active and together with the
$B_{1u}$ mode gives the $A_u$ representation, $8A_{u}+9B_{1u}\rightarrow 17A_{u}$. A possible
twinning in the $b-c$ plane cannot be a reason for leaking the $b-c$ phonons to the $a$ axis and
vice versa. Alternatively, in the lowest $C_i^1$ symmetry all the phonons mix together, as seen in
Fig.~\ref{fig:DescentInSymmetry}. In the $Pb11$ symmetry, Raman $A_{g}$ modes become visible in the
$bc$ plane and together with the IR $B_u$ phonons form the $A'$ representation,
$12A_{g}+18B_{u}\rightarrow 30A'$.  Similar, Raman $B_g$ modes can be observed in the $a$ axis
response together with IR $A_u$ modes, $12B_{g}+18A_{u}\rightarrow 30A''$. An important difference
between $P\overline{1}$ and $Pb11$ symmetries is that in $P\overline{1}$ the same modes can be
observed in all three polarizations, while in $Pb11$ $a$ eigenmodes on one side and $b$ and $c$
eigenmodes on the other side remain separate.

At room temperature we observed 9, 9, and 7 phonons for the $a$, $b$, and $c$ axes, respectively
(see Fig.~\ref{fig:ReflectivityABC}). The TO phonon frequencies $\omega$, broadening $\Gamma$ and
optical strength $S$ are presented in Tables~\ref{table1} and \ref{table2}. Our observation
confirms the orthorhombic $Pbnm$ symmetry of the crystal lattice at room temperature. At lower
temperatures, as far as the crystal structure is concerned, we can conclude from
Fig.~\ref{fig:NewModes} that the $Pbnm$ symmetry is transformed to a lower one below 200 K and it
is restored again below 75 K. Evolution of the optical strength of the new phonons with temperature
indicates that the transition at 200 K. Surprisingly it is restored again below 75 K. The
temperature evolution of the optical strength of the new phonons indicates that the phase
transition at 200 K has a second order nature and that the one at 75 K has a first order nature.
Shown in Table \ref{table2} are the phonon mode frequencies and strengths observed in the $c$ axis
spectrum at 75 K. The lines active also in the high temperature phase are in bold face and assigned
to $B_{2u}$, those appearing only in the intermediate temperature phase are labelled by $A$.
Although much more new phonons are seen for all three axes, only the activated $c$-axis modes,
having a counterpart in the other axes, are shown in the table. First, one can see that the
original $c$-axis phonon lines give rise to phonon absorptions along the $b$ axis. The reverse is
also valid. The phonons from the $b$ axis `leak' in the intermediate phase to the $c$ axis, which
is shown in Fig.~\ref{fig:NewModes}. This agrees with all three symmetries under consideration.
However, there is a number of newly activated absorption peaks (199, 350, and 522 cm$^{-1}$) in the
$b$ and $c$ polarizations, which do not have parents in other polarizations. The fact that these
new modes are exactly at the same frequency in both $b$ and $c$ polarizations - as shown in the
right panels of Fig.~\ref{fig:NewModes} - exclude the possibility of the doubling of the unit cell
as their origin. The appearance of new modes is forbidden in the $P2_1/a$ symmetry (see also
Fig.~\ref{fig:DescentInSymmetry}). This observation rules out the $P2_1/a$ symmetry for the
intermediate phase. Because these modes are not seen in any polarization above 200 K or below 75 K,
we can ascribe these phonons either to the previously silent mode $A_u$ activated in the
$P\overline{1}$ symmetry or to the high temperature Raman modes activated in $Pb11$ symmetry. We
did not observe any correlation of phonons between $b$ and $c$ on one side and $a$ on the other
side, apart from the modes already present both in $a$ and $c$ polarizations above 200 K, 333 and
420 cm$^{-1}$ modes in Table~\ref{table2}. This could imply that the triclinic distortions are
extremely small, in particular for the $a$ axis. However, the $Pb11$ symmetry gives a more natural
explanation, as it keeps the $a$ and $bc$ modes apart consistent with the observation. All in all,
these facts draw us to the conclusion that the crystal structure of YVO$_3$ in the intermediate
phase most likely has a monoclinic $Pb11$ symmetry. A further test for this conclusion could be a
comparison to Raman measurements, when data becomes available.

\begin{figure}
\includegraphics[clip=true,width=0.47\textwidth]{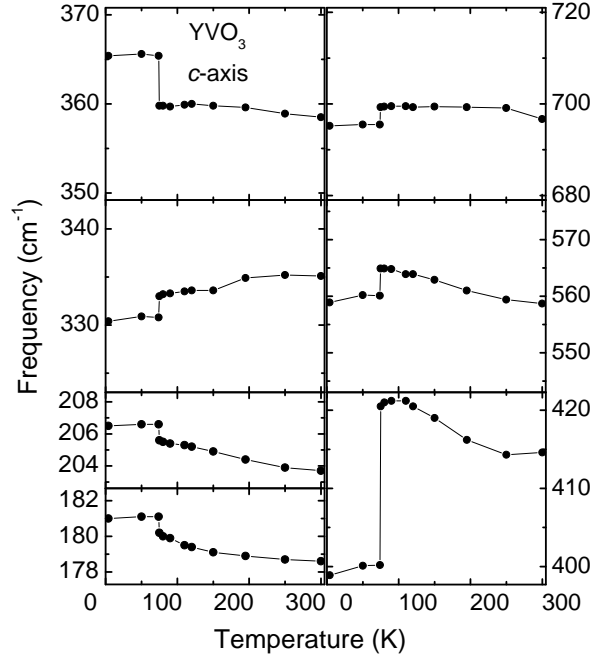}
\caption{\label{fig:Conductivity_T_dependence}Temperature dependence of the TO frequencies for the
$c$-axis $B_{2u}$ phonon modes. For comparison purposes, vertical scale in each square comprises
6\% of the central frequency.}
\end{figure}

Finally, we remark on the relation between the electronic and  crystal structure changes. From
Fig.~\ref{fig:Conductivity_T_dependence} one can see that the main modes evolve smoothly through
the second order phase transition, where the appearance of the orbital order is
reported\cite{Blake-PRL-2001,Blake-PRB-2002}. However, modes change their frequency substantially
at the first order phase transition changes some, where spins and orbitals switch their ordering.
The frequency change is found to be particularly strong for the modes around 400 cm$^{-1}$. From
lattice dynamical calculations\cite{Smirnova-PhysicaB-1999} for the isostructural compound
LaMnO$_3$ we conclude that the 400 cm$^{-1}$ modes mainly involve vanadium-oxygen bond vibrations.
Therefore, it seems that the main structural changes at the first order transition involve a
distortion of the VO$_6$ octahedra.


\section{Electronic structure}

\begin{figure}
\includegraphics[clip=true, width=0.46\textwidth]{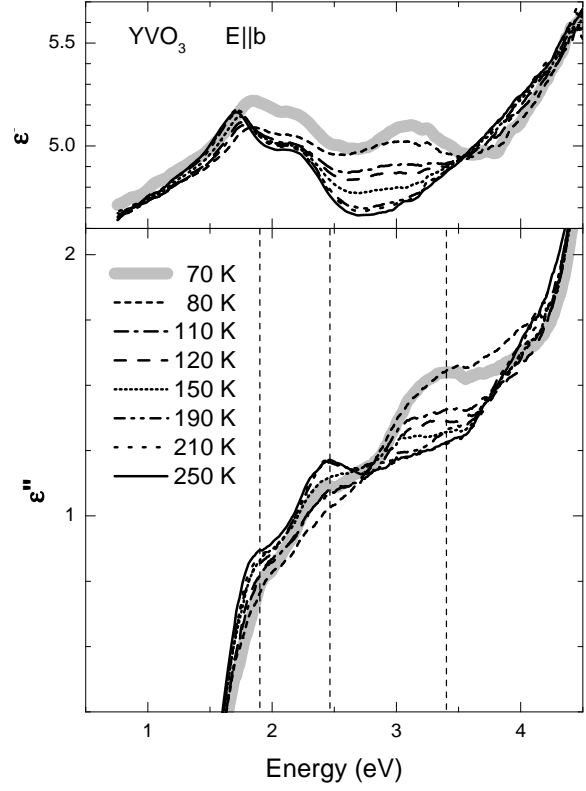}
\caption{Temperature dependence of the real and imaginary part of the b-axis dielectric function
for YVO$_3$. The vertical lines indicate locations of the band maxima.} \label{fig:EpsilonSigma}
\end{figure}

The yttrium Y$^{3+}$ and vanadium V$^{3+}$ ions in YVO$_3$  both carry the formal charge $3+$.
Yttrium has a complete shell, and it can be excluded from the consideration for the electronic
transitions. Vanadium has two $3d$ electrons in the outer shell. We can thus expect two types of
optical transitions in the visible range: $d-d$ transitions between vanadium ions with the
characteristic energy of the Mott-Hubbard gap (Mott-Hubbard (MH) transitions), and the
charge-transfer (CT) transitions from the oxygen $2p$-band to the free states in the vanadium
$3d$-band.

\begin{figure}
\includegraphics[clip=true, width=0.40\textwidth]{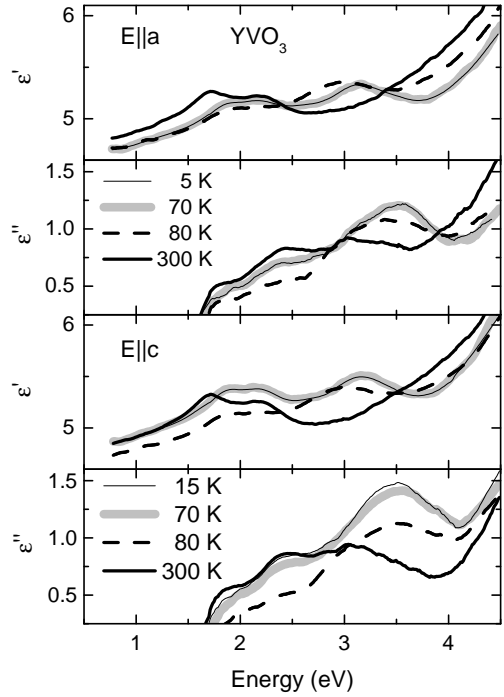}
\caption{The real and imaginary part of the dielectric function for the $a$- and $c$-axes of
YVO$_3$.} \label{fig:EpsilonAC}
\end{figure}

Shown in Figs.~\ref{fig:EpsilonSigma} and \ref{fig:EpsilonAC} are the real and imaginary parts of
the YVO$_3$ dielectric function, $\varepsilon=\varepsilon'+i\varepsilon''$, measured at various
temperatures. The optical response at low temperatures is determined by three bands with their
maxima located at 1.8, 2.4, and 3.3 eV, followed by a sharp upturn above 4 eV. The optical gap is
found at 1.6 eV. It is quite remarkable to observe that the intensity of some bands, located in the
region of a few electron-volts, changes by as much as 50\% for temperature variations as low as
$k_BT \leq 26$ meV.

Recently Miyasaka {\it et. al.}\cite{Miyasaka-JPSJ-2002} reported reflectivity measurements on
YVO$_3$, where the optical conductivity, obtained through the Kramers-Kronig relations, revealed
only two peaks. Also the temperature dependence and the reported anisotropy of the conductivity
were different. Since $\varepsilon''$ is approximately five times smaller than $\varepsilon'$ in
the visible part of the spectrum, the reflectivity is mainly determined by the real part of
$\varepsilon$. The $d$-band structure in $\varepsilon'$ and $\varepsilon''$ gives rise to the
corresponding reflectivity variations of only few percent and the temperature induced changes of
the reflectivity are around 1\% or less. This together with the close vicinity of the strong charge
transfer transitions place heavy demands on the accuracy of the measurements. A small error can
substantially affect an interpretation of the electronic spectrum. We used the ellipsometric
technique, which determined $\varepsilon'$ and $\varepsilon''$ directly at each particular photon
energy and excluded the necessity of reference measurements. In addition we took special
precautions to perform measurements in ultra-high vacuum in order to avoid condensation on the
surface at low temperatures. We trust that these measures minimize a possible error.

Let us first discuss the origin of the bands. Following the general classification
scheme\cite{Zaanen-PRL-1985}, YVO$_3$ is a Mott-Hubbard insulator. The lowest excitations are
$d^2_i d^2_j\rightarrow d^3_i d^1_j$ transitions between different vanadium ions, $i$ and $j$. The
upturn above 4 eV is due to charge transfer excitations. This general picture follows from our
LSDA+$U$ calculations and agrees with previous room temperature
measurements\cite{Kasuya-PRB-1993,Arima-PRB-1993}. The fine structure of the MH band observed here
has not been reported previously and will be discussed using ligand field theoretical
considerations and LSDA+U calculations.

\begin{figure}
\includegraphics[clip=true,width=0.40\textwidth]{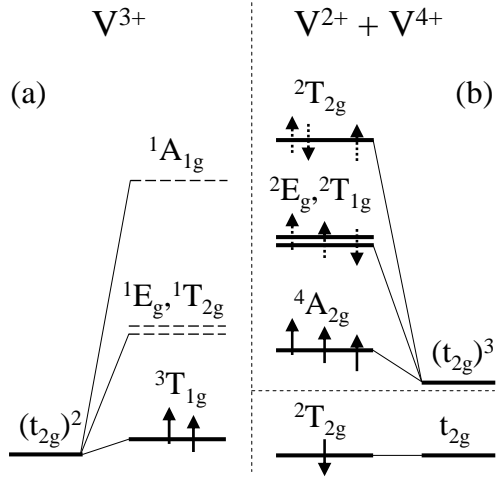}
\caption{Energy diagram for (a) the ground state of the vanadium ions and (b) excited states of two
vanadium sites.} \label{fig:EnergyDia}
\end{figure}

Because the orthorhombic distortion of the oxygen octahedron is small, we can neglect any on-site
d-d excitations. For the same reason we can, considering the energy diagram, restrict ourselves in
the first approximation to the O$_h$ symmetry on the vanadium site. The $e_g$ orbitals can be
excluded, because due to the large crystal field splitting, $10Dq$, the mixing between $e_g$ and
$t_{2g}$ orbitals is expected to be small. Thus, in the strong field approximation the two $d$
electrons occupy $t_{2g}$ orbitals.

If we take into account the Coulumb interaction, the $t_{2g}$ level is split into $^3T_{1g}$,
$^1E_{g}$, $^1T_{2g}$, and $^1A_{1g}$ levels, as shown in Fig.~\ref{fig:EnergyDia}~(a). The
$^1E_{g}$ and $^1T_{2g}$ levels are degenerate in the O$_h$ symmetry. In the ground state the $d$
electrons occupy the lowest energy S=1 state, $^3T_{1g}$, in agreement with the Hund's rule. The MH
excitations occur from the initial state with two vanadium in the $d^2$ state to the final state
with the first atom in the $d^1$ (V$^{4+}$) and the second in the $d^3$ (V$^{2+}$) states. In the
final state the wave function of the single electron follows obviously only one irreducable
representation $^2T_{2g}$, while the $d^3$ state splits into four levels, $^4A_{2g}$, $^2E_{g}$,
$^2T_{1g}$, and $^2T_{2g}$, with $^2E_{g}$ and $^2T_{1g}$ again being degenerate. The energy
diagram of the final state is depicted in Fig.~\ref{fig:EnergyDia}~(b). The three observed bands in
Fig.~\ref{fig:EpsilonSigma} correspond to transitions to these three energy levels. According to
the measurements, these bands span the energy range of 1.5 eV from 1.8 eV to 3.3 eV. With the
crystal field splitting of $10Dq \approx 2$ eV, our assumption of a small admixture of $e_g$
orbitals is justified.

\begin{figure}
\includegraphics[clip=true,width=0.47\textwidth]{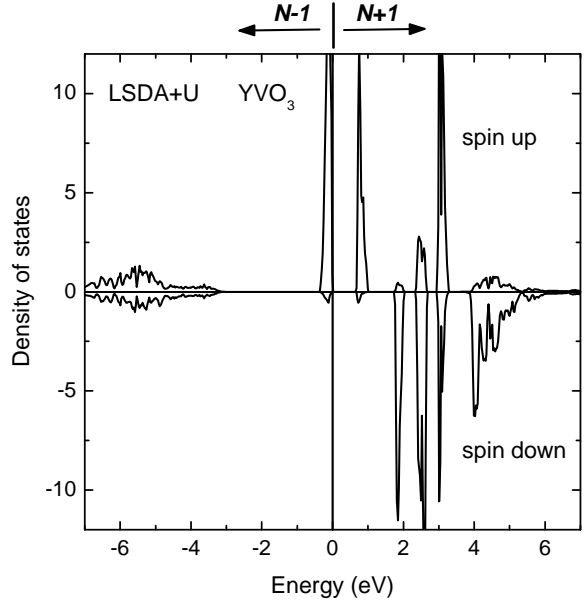}
\caption{Vanadium $3d$ partial density of states for spin-up and spin-down electrons, obtained in
local spin-density approximation with the Hubbard potential $U$ taken into account. $N-1$ and $N+1$
correspond to the electron removal and electron addition states, respectively.} \label{fig:LSDA-U}
\end{figure}

The picture obtained with the ligand field theory is in good agreement with our LSDA+$U$
calculations. The LSDA+$U$ band structure calculations has been performed in the LMTO calculation
scheme\cite{Andersen-PRB-1975,Anisimov-PRB-1991}. Vanadium $3d$ partial density of states were
obtained in the LDA+U calculation with $U=3.4$ eV and $J=0.85$ eV for the low temperature crystal
and magnetic structure of YVO3. The results are consistent with previous calculations of the
electronic structure\cite{Sawada-PRB-1998}. As can be seen from the electron density of states in
Fig.~\ref{fig:LSDA-U}, the highest occupied states are from a narrow $d$ band, with both electrons
in a spin-up state. The AFM ordered neighboring vanadium ion should be accordingly in a spin-down
state. The lowest unoccupied band corresponds to the addition of an electron with spin up. This
agrees with the lowest excited $^4A_{2g}$ state, which indeed puts electrons in the high-spin state
$S=3/2$. Two consequent excitation bands are formed by transfer of spin-down electrons, again in
agreement with the low $S=1/2$ spin states in the $^2E_{g}$, $^2T_{1g}$, and $^2T_{2g}$
representations found from the ligand field theory. Higher electron-addition bands with equal
spin-up/spin-down populations have an $e_g$ origin. From the energy distribution of the bands in
Fig.~\ref{fig:LSDA-U} one can see that YVO$_3$ is indeed a Mott-Hubbard
insulator\cite{Zaanen-PRL-1985}. The lowest excitations are $d-d$ transitions, $d^2d^2\rightarrow
d^3d^1$. The charge-transfer transitions, $d^2\rightarrow d^3\overline{L}$ occur at higher energies
than the $d-d$ transitions.

\begin{figure}
\includegraphics[clip=true,width=0.47\textwidth]{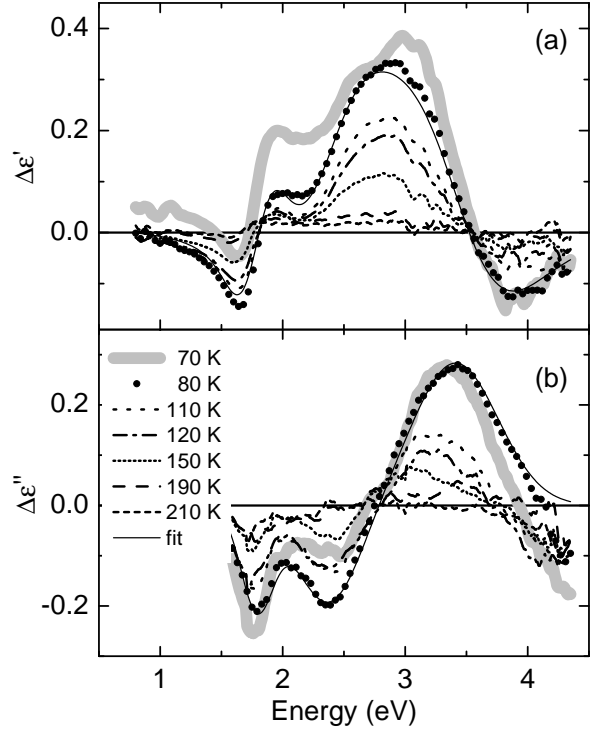}
\caption{\label{fig:Diffee}The dielectric functions from Fig.~\ref{fig:EpsilonSigma}, the real and
imaginary parts, with subtracted 250 K data. The fit curve shows a simultaneous fit to the real and
imaginary part for the 80 K data.}
\end{figure}

\begin{figure}
\includegraphics[clip=true,width=0.47\textwidth]{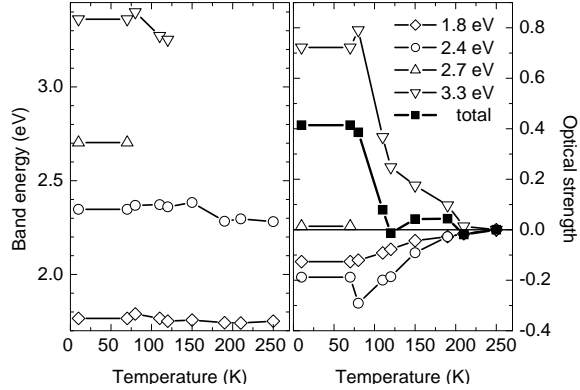}
\caption{\label{fig:3bands}Temperature dependence of the bands position (left) and their optical
strength together with the total spectral weight of the three bands (right).}
\end{figure}

The temperature dependence of the optical response of YVO$_3$ is shown in
Figs.~\ref{fig:EpsilonSigma} and \ref{fig:EpsilonAC}. There is only small T-dependence between 70 K
and 6 K, as can be seen in Fig.~\ref{fig:EpsilonAC}, and the corresponding data are further omitted
from the consideration. To clarify the temperature dependence of the MH bands,
Fig.~\ref{fig:Diffee} shows a differential dielectric function for $E||b$, where the 250 K data are
subtracted from the curves in Fig.~\ref{fig:EpsilonSigma}. In order to quantify temperature
variations of the optical strength of the bands, we fitted simultaneously the real and imaginary
parts. The differential dielectric function turned out to be well fitted with three gaussians in
the imaginary part and with its Hilbert transform, Dawson function, in the real part. The fitted
curves for 80 K are shown along with the experimental data in Fig.~\ref{fig:Diffee}. The full
dielectric function shown in Fig.~\ref{fig:EpsilonSigma} was also fitted in the same way, but an
additional fourth gaussian was necessary to account for the CT band. In this way the position of
the bands was confirmed to be temperature independent. However, it was not possible to determine
accurately the absolute optical strength of the MH bands due to an additional non-gaussian
contribution. It was also impossible to determine the position of the 3.3 eV band at high
temperatures because of its vanishing presence in the spectra. The band positions and optical
strengths obtained from the fits are plotted in Fig.~\ref{fig:3bands}. The dielectric function
reveals minimal variations between room temperature and the second order structural phase
transition at 200 K. Below 200 K a strong growth of the 3.3 eV band is observed, which is
accompanied by the decrease of the 1.8 and 2.4 eV bands. The total change of the optical strength
of these MH bands remains virtually zero down to the N\'eel temperature, $T_N=116$ K. In the
magnetically ordered state the 3.3 eV band acquires additional strength. The additional spectral
weight transfer should come from higher energies, most probably from the $e_g$ states, as the
decrease of the low energy bands cannot account for it. Indeed, the lowest energy state in the
$t_{2g}^2 e_g$ configuration is obtained by adding an $e_g$ electron to the ground state
$^3T_{1g}$. Four irreducible representations, $^4T_{1g}$, $^4T_{2g}$, $^2T_{1g}$, and $^2T_{2g}$
can be formed from this configuration, where the latter representation descends also from the
$t_{2g}^3$ configuration. Antiferromagnetic order favors transitions to the low-spin states and
thus gives additional optical strength to the $^2T_{2g}$ state.

At the first order phase transition, the optical strength of the 2.4 and 3.3 eV bands experiences
jumps in opposite directions. Additionally, as can be seen in Fig.~\ref{fig:Diffee}~(a), a new
shoulder emerges at around 2.7 eV. This appears in Fig.~\ref{fig:EpsilonSigma} as an extended flat
region, which becomes visible probably due to the narrowing of the 3.3 eV band. However, its
optical strength is very small.

The optical spectrum begins to change below the second order phase transition at 200 K. These
changes can not be directly related to the lower symmetry as the 3.3 eV band does not disappear in
the low temperature orthorhombic phase. The reason should rather come from some electronic
correlations and associated distortions of the VO$_6$ octahedrons. Ren {\it et
al.}\cite{Ren-Nature-1998} suggested that the spin order in YVO$_3$ is accompanied by an orbital
order. The orbital order is reported to appear below the 200 K phase
transition\cite{Blake-PRL-2001, Blake-PRB-2002}. In the proposed picture for the orbital ordering,
the $xy$-orbitals are always occupied by one of the two $d$ electrons, while the occupation of the
$xz$ and $yz$ orbitals alternates in an AFM order. The orbital order was found to be different
above and below 77 K in agreement with the different magnetic order. According to Ren {\it et
al.}\cite{Ren-Nature-1998}, there is a $C$-type magnetic and $G$-type orbital order above 77 K, and
alternatively $G$-type magnetic and $C$-type orbital order below 77 K.

We now turn to the discussion of how orbital and magnetic ordering influences the optical
conductivity. We consider a model, where the optical conductivity is proportional to the electron
transition probability between two neighboring vanadium sites. There are in total nine independent
eigenfunctions in the $^3T_{1g}$ representation. At room temperature the electron transition
probability can be calculated by averaging over arbitrary orientations of spin and orbits on two
ions, i.e. by averaging over all linear combinations of these 9 wave functions. Magnetic or orbital
order limits the choice of the initial wave functions. At lower temperatures we can expect that the
electron transition from the S=1 state on the first ion to the S=3/2 state on the other ion becomes
less probable due to the antiferromagnetic order. Thus, the intensity of the lowest band should
decrease. As can be seen from Fig.~\ref{fig:EpsilonSigma} and Fig.~\ref{fig:Diffee}, the optical
strength of the lowest band indeed decreases together with temperature. By using a more rigorous
analysis of the transition probabilities, we can evaluate if a particular order is consistent with
the observed temperature dependence for $\varepsilon(\omega)$. In this analysis we neglect the
tilting of octahedrons, and assume that the transition integrals between $d$-orbitals on the
neighbor vanadium ions are the same for all directions and for all $d$ orbitals. The conductivity
is assumed to be determined by the probabilities to create a pair consisting of an occupied orbital
on one site and an empty orbital on the neighboring site, with the constraint that the transition
is allowed. The intersite $d-d$ transitions are mediated by the $p$ orbitals of the oxygen ions
between the vanadium sites. At room temperature all $^3T_{1g}$ states are occupied randomly. At
lower temperatures we used the magnetic and orbital order, suggested above\cite{Ren-Nature-1998}.
If we assume the transition integral from completely occupied to completely empty orbital as unity,
then the transition probabilities at room temperature are found to be 0.49, 0.84, and 0.44 for the
$^4A_{2g}$, $^2E_{g} + ^2T_{1g}$, and $^2T_{2g}$ final states, respectively. We also calculated the
transition probabilities for AFM spin order and for AFM spin order accompanied by the
aforementioned orbital order (see Table~\ref{table4}). Both simple AFM-SO and AFM-SO+OO lead to a
decrease of the lowest band and an increase of the intensity of the highest band, consistent with
the observations.

However, none of the considered orderings gives the correct  sign for the intensity change for the
middle band, {\it i.e.} a decrease of strength with decreasing temperature. Moreover, we found that
no combination of completely polarized states ($S_z=1$) can give the desired intensity decrease for
the $^2E_{g} + ^2T_{1g}$ band. The same conclusions hold if the on-site symmetry is lowered to
tetragonal. If some addition of the $S_z=0$ initial wave function is allowed, as suggested by
neutron measurements\cite{Blake-PRB-2002}, a whole number of possible orbitally ordered structures
becomes allowed. Although different orbital orderings can indeed explain the observed intensity
changes, it is not possible to give a preference to a particular order\cite{Note-1}.


\section{Magneto-optical Kerr effect}

\begin{figure}
\includegraphics[clip=true,width=0.47\textwidth]{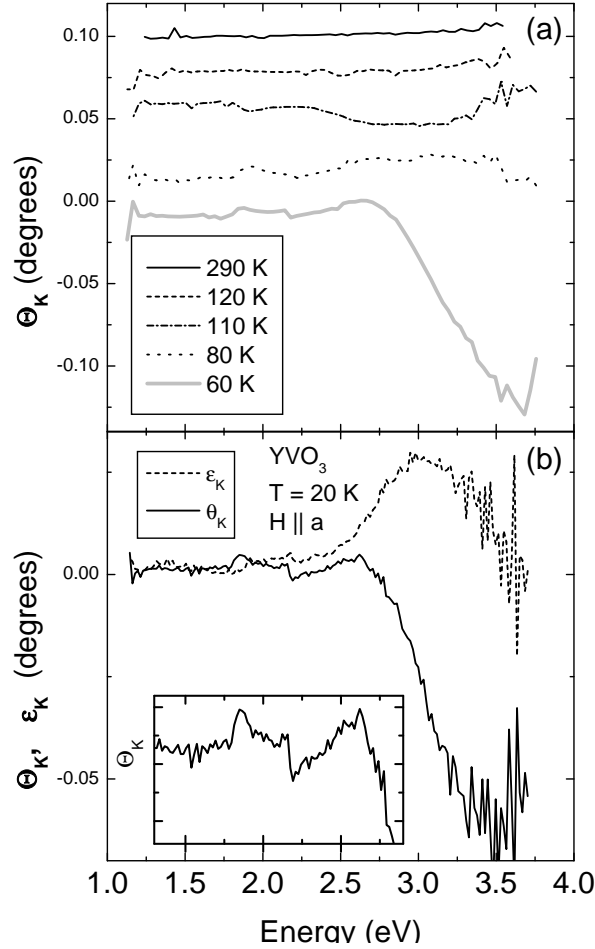}
\caption{\label{fig:Kerr}(a) Temperature dependence of the Kerr rotation measured in the polar
geometry with $H||a$. (b) The Kerr rotation $\Theta_K$ together with the Kerr ellipticity
$\varepsilon_K$ in the low temperature phase. (Insert) Enlarged view of the resonances at 1.8 and
2.2 eV.}
\end{figure}

The energy diagram shown in Fig.~\ref{fig:EnergyDia} was derived assuming a local cubic symmetry on
the vanadium site. In reality, one axis of the octahedron is different from the other two. This
tetragonal type distortion additionally splits the multiplet structure, which partially lifts the
degeneracy of the ground and excited states. On top of that, a spin-orbital coupling provides
segregation of the states by the quantum number $M_J$. The latter gives rise to the magneto-optical
Kerr effect (MOKE). Temperature dependence of the Kerr rotation spectrum $\Theta_K(E)$ is shown in
Fig.~\ref{fig:Kerr}. The curves are shifted for clarity. The measurements were performed in the
polar geometry at nearly normal angle of incidence, with magnetic field of 0.1-0.2 T parallel to
the $a$ axis, $H||a$. In the other field orientations, the magnetic moment was too small for the
Kerr rotation to be detected. No rotation of the polarization plane was observed above the N\'eel
temperature within our experimental error. Below the N\'eel temperature but above the first order
phase transition, only a weak structure of the order of $\Theta_K\approx0.01^\circ$ can be seen.
Below 75 K a remarkable peak as large as $\Theta_K=0.1^\circ$ is observed. In total, there are
three resonance structures. Enlarged in the insert of Fig.~\ref{fig:Kerr} are two surprisingly
narrow dispersion type resonances at 1.8 and 2.2 eV that have positions approximately corresponding
to the onset of the two first MH bands. Their intensity, though rather small $\sim0.01^{\circ}$, is
well reproduced on different samples. The third broad peak appears to have maximum at 3.5 eV, which
is also confirmed by the zero crossing of the Kerr ellipticity $\varepsilon_K(E)$. This band lies
below the charge-transfer onset and corresponds to the $d-d$ transitions. The maximum rotation in
this band amounts to $\sim0.1^{\circ}$. Such a large magneto-optical effect seems to be quite
surprising, because from the electrodynamics point of view the time-reversal symmetry, although
broken on the microscopic scale, is usually restored on the macroscopic scale in antiferromagnetic
compounds. If this indeed were the case in YVO$_3$, the Kerr rotation would mainly result from the
small $0.2^{\circ}$ spin canting and would be as minuscule as the net magnetic moment is. For
comparison, the saturation magnetic moment in Ni is $0.61\mu_B$ per atom, i.e. approximately 60
times larger that the net moment of 0.01 Bohr magneton per vanadium atom in YVO$_3$. Still the Kerr
rotation in Ni is quite comparable to those of the low temperature phase of YVO$_3$ and lies
between $0.1^{\circ}$ and $0.2^{\circ}$ in the visible range.

Antiferromagnetic compounds can have large magneto-optical effects, as was shown
recently\cite{Solovyev-PRL-1996,Sandratskii-PRL-1996}, if the local spin and orbital magnetic
moments are not collinear. In particular, an orbital magnetic moment does not vanish in
transition-metal perovskites with the $Pbnm$ structure due to the tilting of the octahedra
\cite{Solovyev-PRB-1997}. We can find non-vanishing ferromagnetic components of the net orbital
magnetic moment $\mathbf{M}_L$ from symmetry considerations, following the guideline and notations
from Ref.~\onlinecite{Solovyev-PRB-1997}. There are four inequivalent vanadium sites $i=1,2,3,4$ in
the YVO$_3$ unit cell. These sites can be generated by four symmetry transformations, unity
operation $E$, and three screw axes, $S_{2x}$ with the shift ($\frac{1}{2}$,0,0) and coordinates
x,$\frac{1}{4}$,0; $S_{2y}$(0,$\frac{1}{2}$,0) $\frac{1}{4}$,y,$\frac{1}{4}$; and
$S_{2z}$(0,0,$\frac{1}{2}$) 0,0,z. Here $x$,$y$, and $z$ are parallel to the crystallographic axes
$a$,$b$,and $c$, respectively. The local relation between the orbital magnetic moment
$\mathbf{M}^i_L$ and spin magnetization direction
$\mathbf{e}=(\cos\theta\sin\phi,\sin\theta\sin\phi,\cos\theta)$ on each vanadium site $i$ can
generally be written as $M^i_{L\alpha}=\mathcal{L}^{i}_{\alpha\beta}e_{\beta}$, where the matrix
$\mathcal{L}^i_{\alpha\beta}$ determines the orbital magnetic moment induced by the spin-orbit
interaction. We apply the mentioned symmetry operations to the matrix $\mathcal{L}^i_{\alpha\beta}$
to obtain averaged matrices $\mathcal{L}_{\alpha\beta}=\sum^4_{i=1} \mathcal{L}^i_{\alpha\beta}$
for the $C$- and $G$-type spin ordering, respectively:
\[
\mathcal{L}^C = 4\left(
\begin{array}{ccc}
0 & \mathcal{L}^1_{xy} & 0\\
\mathcal{L}^1_{yx} & 0 & 0\\
0 & 0 & 0
\end{array}
\right), \,\,\, \mathcal{L}^G = 4\left(
\begin{array}{ccc}
0 & 0 & \mathcal{L}^1_{xz}\\
0 & 0 & 0\\
\mathcal{L}^1_{zx} & 0 & 0\\
\end{array}\right)
\]
The corresponding net orbital magnetic moments are
\[
\begin{array}{ll}
C: &
\mathbf{M}^C_L=4(\mathcal{L}^1_{xy}\sin\theta\sin\phi,\mathcal{L}^1_{xy}\sin\theta\cos\phi,0),\\
G: &
\mathbf{M}^G_L=4(\mathcal{L}^1_{xz}\cos\theta,0,\mathcal{L}^1_{zx}\sin\theta\cos\phi).
\end{array}
\]
The orbital magnetic moment $\mathbf{M}_L$ has the same symmetry as the gyration vector
$\mathbf{g}$\cite{Uspenskii-PRB-1996}, which determines the antisymmetric part of the dielectric
tensor, $\varepsilon^a_{\alpha\beta}=ie_{\alpha\beta\gamma}g_{\gamma}$, which is responsible for
mageto-optical effects\cite{LandushitzVIII}. Here $e_{\alpha\beta\gamma}$ is the unit antisymmetric
tensor. In the polar measurement geometry discussed here, components of the vector $g_{\alpha}$ and
correspondingly components of $\mathbf{M}_L$ define the orientation of magnetic field $H$, at which
the observation of Kerr effect is possible.

For the G-type spin order, the structure of $\mathbf{M}^G_L$ suggests that the polar Kerr effect
can exist for $H||a$, if spins are parallel to the $c$ axis, and for $H||c$, if spins are aligned
along the $a$ axis. If the spins are along the $b$ axis, $\mathbf{M}^G_L$ and the corresponding
Kerr effect are zero. This does not exclude the possibility for the second, weaker, type of Kerr
effect, which is directly related to the spin canting. It should be emphasized here that the Kerr
rotation due to the orbital magnetic moments can exist even if spins are exactly
antiferromagnetically aligned\cite{Solovyev-PRB-1997}. Our observation of the substantial
magneto-optical effect only for $H||a$ shows that spins in the low temperature phase of YVO$_3$ are
aligned along the $c$ axis. This agrees with the neutron scattering
measurements\cite{Blake-PRB-2002}. The weak structure at 1.8 and 2.2 eV is most probably the spin
canting effect.

The C-type spin order as seen from the $\mathbf{M}^C_L$ angular dependence allows a magneto-optical
activity, only if spins are aligned in the $ab$ plane. The small magnitude of the experimentally
observed MOKE above 80 K compared to the low temperature phase suggests that the spins are aligned
along the $c$ axis. In this case, $\mathbf{M}^C_L$ is zero for any orientation of $H$, and Kerr
effect substantially decreases, in agreement with the our experimental observations. This result
contradicts the neutron data, where a substantial spin component along the $b$ axis is
observed\cite{Blake-PRB-2002}. This, however, can reflect another important property of YVO$_3$. As
it was shown in Ref.~\onlinecite{Jo-JPSJ-2003}, the G-type antiferromagnetic orbital order in the
form proposed by Ren {\it et. al.}\cite{Ren-Nature-1998} suppresses the orbital magnetic moment.
This can be the reason, why no substantial Kerr rotation was observed for the C-type spin order.
This conclusion requires that the on-site magnetic moment in the $G$-type orbital order is reduced
from $2\mu_B$ to $1.1\mu_B$\cite{Blake-PRB-2002} not due to the orbital magnetic moment, but due to
a mixture of $S_z=1$ and $S_z=0$ $^3T_{1g}$ states, in agreement with the conclusion obtained in
the previous section.

\section{Conclusions}

The analysis of the YVO$_3$ phonon spectra revealed three different phases. The orthorhombic $Pbnm$
structure exists down to 200 K. At 200 K there is a second order phase transition to a lower
symmetry structure. Orthorhombic $Pbnm$ crystal symmetry is restored through a first order phase
transition below 74 K. The most probable structure in the intermediate phase is the monoclinic
$Pb11$, where the VO$_6$ octahedra form a dimerized chain along the $c$ axis. The triclinic
$P\overline{1}$ symmetry cannot however be completely excluded.

YVO$_3$ is a Mott-Hubbard insulator with the optical gap of 1.6 eV. There are three optical bands
with energies 1.8, 2.4, and 3.3 eV. We identify these bands with the inter-vanadium transitions
from the $^3T_{1g}$ ground state to the $^4A_{2g}$, $^2E_{g} + ^2T_{1g}$, and $^2T_{2g}$ states.
These $d-d$ excitations are followed by the charge-transfer transitions at 4 eV. The band gap
structure is in agreement with the LSDA+$U$ calculations. The $d$ bands revealed strong temperature
dependence. Using a model, which includes only the nearest neighbor transitions, we found that the
observed temperature dependence disagrees with the simple antiferromagnetic spin ordering. This
indicates that an orbital order is responsible for the band temperature variations. However, the
commonly considered $|xy,xz|/|xy,yz|$ orbital order seems also inconsistent with our data, if we
assume a fully polarized on-site state.

We observed two types of magneto-optical effect. Small net magnetic moment due to the uncompensated
spin moments induces the weak Kerr rotation of the order of $\sim0.01^{\circ}$. The effect exists
in all phases below the magnetic ordering temperature and for all active optical transitions. A
strong Kerr effect appears as a consequence of the low crystallographic symmetry and ferromagnetic
ordering of orbital magnetic moments. The Kerr rotation of the order of $\sim0.1^{\circ}$ was found
in the low temperature phase for the highest $d-d$ transition band. It follows from the symmetry
considerations that the spin magnetic moment lays along the $c$ axis in both magnetic structures.
However, this conclusion for the C-type AFM spin arrangement can be affected by the accompanying
orbital order.

\acknowledgments

We acknowledge useful discussion with G. R. Blake. We thank H.J. Bron for his help with sample
preparation. This research was supported by the Netherlands Foundation for Fundamental Research on
Matter (FOM) with financial aid from the Nederlandse Organisatie voor Wetenschappelijk Onderzoek
(NWO).
\bibliography{yvo3paper}

\begin{thebibliography}{27}
\bibitem[{Not()}]{Note-1}
 (????), \bibinfo{note}{for example, if we maximize the spin magnetic moment,
  we obtain that the $|xz,yz|$ orbitals should be always occupied on all
  vanadium sites. This gives correct intensity changes for the all bands and
  all axes.}
\end{thebibliography}



%


\begin{table}[b]
\caption{TO phonon modes for the $a$ and $b$ axes at room temperature.} \label{table1}
\begin{ruledtabular}
\begin{tabular*}{\hsize}{@{\extracolsep{0pt}}cdd@{\extracolsep{0ptplus1fil}}c@{\extracolsep{0pt}}cdd@{\extracolsep{0pt}}}
\multicolumn{3}{c}{$B_{1u}$ (a)} & \ \ & \multicolumn{3}{c}{$B_{3u}$ (b)}\\
\colrule
$\omega$, cm$^{-1}$ & \multicolumn{1}{c}{$\gamma$, cm$^{-1}$} & \multicolumn{1}{c}{$S_a$} & \ \ & $\omega$, cm$^{-1}$ & \multicolumn{1}{c}{$\gamma$, cm$^{-1}$} & \multicolumn{1}{c}{$S_b$} \\
\colrule
162.2 & 3.2 & 0.08 & \, \, & 144.3 &  15.5  &  0.73 \\
211.6 & 8.5 & 5.4 & \, \, & 237.4 &   5.5  &  2.0 \\
328.5 & 9.9 & 1.6 & \, \, & 280.1 & 8.4  &  0.64 \\
359.2 & 15.0 & 0.56 & \, \, & 318.7 &   7.8  & 0.35 \\
420.8 & 10.4 & 1.4 & \, \, & 374.5 &   6.3  &  0.22 \\
446.8 & 17.3 & 2.6 & \ \ & 390.7 &  16.5  &  4.21 \\
494.5 & 10.0 & 0.005 & \ \ & 471.7 &  13.3  &  1.2 \\
517.5 & 15.5 & 0.06 & \ \ & 543.7 &   9.8  &  0.86 \\
572.4 & 22.6 & 0.91 & \ \ & 559.8 &  17.5  &  0.21 \\
\end{tabular*}
\end{ruledtabular}
\end{table}

\begin{table}[b]
\caption{c-Axis phonons at 75 K. $B_{2u}$ modes originate from the room temperature phase. New
vibrational modes, appearing below 200 K, are designated as $A$} \label{table2}
\begin{ruledtabular}
\begin{tabular}{lcddd}
Mode & \mbox{Frequency} & \multicolumn{1}{c}{$S_a$} & \multicolumn{1}{c}{$S_b$} & \multicolumn{1}{c}{$S_c$} \\
\colrule
$B_{2u}$ & {\bf 180.2}  & -     & -     & 7     \\
$A$      & 199          & -     & 0.037 & 0.008 \\
$B_{2u}$ & {\bf 205.6}  & -     & 0.006 & 0.78  \\
$B_{2u}$ & {\bf 333.0}  & 0.19  & 0.008 & 2.9   \\
$A$      & 350          & -     & 0.67  & 0.003 \\
$B_{2u}$ & {\bf 359.8}  & -     & 0.02  & 0.37  \\
$B_{2u}$ & {\bf 420.5}  & 1.6   & 0.076 & 1.6   \\
$A$      & 522          & -     & 0.024 & 0.018 \\
$B_{2u}$ & {\bf 564.9}  & -     & 0.014 & 1.2   \\
$B_{2u}$ & {\bf 699.2}  & -     & -     & 0.002 \\

\end{tabular}
\end{ruledtabular}
\end{table}

\begin{table}[b]
\caption{Possible symmetries and symmetry operations in the intermediate phase ($75 K < T < 200
K$).} \label{table3}
\begin{ruledtabular}
\begin{tabular*}{\hsize}{clllll}
N   & \multicolumn{1}{l}{$Pnma$}                                 & \multicolumn{1}{l}{$Pbnm$}                               & \multicolumn{1}{l}{$P2_1/a$ \, } & \multicolumn{1}{l}{$P\overline{1}$} & \multicolumn{1}{l}{$Pb11$} \\
\colrule
    & \multicolumn{1}{l}{$D^{16}_{2h}$}                          & \multicolumn{1}{l}{$D^{16}_{2h}$}                        & \multicolumn{1}{l}{$C_{2h}^5$} & \multicolumn{1}{l}{$C_i^1$ \, } & \multicolumn{1}{l}{$C^2_s$} \\
\colrule
(1) & 1                                                          & 1                                                        & 1            & 1            & 1            \\
(2) & $2(0,0,\frac{1}{2})\,\frac{1}{4},0,z$                      & $2(\frac{1}{2},0,0)\, x,\frac{1}{4},0$                   & $2_1$      & -            & -            \\
(3) & $2(0,\frac{1}{2},0)\,0,y,0$                                & $2(0,0,\frac{1}{2})\,0,0,z$                              & -            & -            & -            \\
(4) & $2(\frac{1}{2},0,0)\,x,\frac{1}{4},\frac{1}{4}$            & $2(0,\frac{1}{2},0)\,\frac{1}{4},y,\frac{1}{4}$          & -            & -            & -            \\
(5) & $\imath$                                                   & $\imath$                                                 & $\imath$     & $\imath$     & -            \\
(6) & $\mathbf{a} \, x,y,\frac{1}{4}$                            & $\mathbf{b} \, \frac{1}{4},y,z$                          & $\mathbf{b}$ & -            & $\mathbf{b}$ \\
(7) & $\mathbf{m} \, x,\frac{1}{4},z$                            & $\mathbf{m} \, x,y,\frac{1}{4}$                          & -            & -            & -            \\
(8) & $\mathbf{n}(0,\frac{1}{2},\frac{1}{2}) \, \frac{1}{4},y,z$ \, & $\mathbf{n}(\frac{1}{2},0,\frac{1}{2})\, x,\frac{1}{4},z$ \, & -            & -            & -            \\
\end{tabular*}
\end{ruledtabular}
\end{table}

\begin{table}[b]
\caption{Calculated transition probabilities at room temperature, in the AFM spin ordered (SO)
state and in the AFM spin and orbital ordered (SO-OO) state.} \label{table4}
\begin{ruledtabular}
\begin{tabular}{lddd}
Order & \multicolumn{1}{c}{$^4A_{2g}$} & \multicolumn{1}{c}{$^2E_{g} + ^2T_{1g}$} & \multicolumn{1}{c}{$^2T_{2g}$} \\
\colrule
Room temperature & 0.49 & 0.84 & 0.44 \\
AFM SO           & 0.30 & 1.48 & 0.89 \\
AFM SO-OO        & 0.33 & 1.67 & 1.0  \\
\end{tabular}
\end{ruledtabular}
\end{table}

\end{document}